\documentclass{article}

\usepackage{mathrsfs}
\usepackage{amsfonts}
\usepackage{amsmath}
\usepackage{amssymb}
\usepackage{graphicx}

\title{Explanation and discovery in aerodynamics}
\author{Gordon McCabe}

\begin{document}

\maketitle

\begin{abstract}

The purpose of this paper is to discuss and clarify the
explanations commonly cited for the aerodynamic lift generated by
a wing, and to then analyse, as a case study of engineering
discovery, the aerodynamic revolutions which have taken place
within Formula 1 in the past 40 years. The paper begins with an
introduction that provides a succinct summary of the mathematics
of fluid mechanics.

\end{abstract}

\section{Introduction}

Aerodynamics is the branch of fluid mechanics which represents air
flows and the forces experienced by a solid body in motion with
respect to an air mass. In most applications of aerodynamics,
Newtonian fluid mechanics is considered to be empirically
adequate, and, accordingly, a body of air is considered to satisfy
either the Euler equations or the Navier-Stokes equations of
Newtonian fluid mechanics. Let us therefore begin with a succinct
exposition of these equations and the mathematical concepts of
fluid mechanics.

The Euler equations are used to represent a so-called `perfect'
fluid, a fluid which is idealised to be free from internal
friction, and which does not experience friction at the boundaries
which are defined for it by solid bodies.\footnote{An `ideal'
fluid is the same thing as a `perfect' fluid.} A perfect Newtonian
fluid is considered to occupy a region of Newtonian space $\Omega
\subset \mathbb{R}^3$, and its behaviour is characterised by a
time-dependent velocity vector field $U_t$, a time-dependent
pressure scalar field $p_t$, and a time-dependent mass density
scalar field $\rho_t$. The internal forces in any continuous
medium, elastic or fluid, are represented by a symmetric
contravariant 2nd-rank tensor field $\sigma$, called the Cauchy
stress tensor, and a perfect fluid is one for which the Cauchy
stress tensor has the form $\sigma = - p g$, where $g$ is the
metric tensor representing the spatial geometry. In component
terms, with the standard Cartesian coordinates $(x_1,x_2,x_3)$ on
Euclidean space, $\sigma^{ij} = -p g^{ij} = -p \delta^{ij}$. Where
$\nabla$ is the covariant derivative of the Levi-Civita connection
corresponding to the Euclidean metric tensor field on Newtonian
space, the Euler equations are as follows:

$$
\rho_t \left [ \frac{\partial U_t}{\partial t} + \nabla_{U_t}U_t
\right ] = - \text{grad} \; p_t + f_t \; .$$ The scalar field
$f_t$ represents the force exerted on the fluid, per unit volume,
by any external force field, such as gravity.\footnote{Given that
a fluid possesses a mass density field $\rho$, it will also be
self-gravitating, in accordance with the Poisson equation:
$\nabla^2 \Phi = - 4 \pi G \rho$. Needless to say, the strength of
a fluid's own gravity is negligible in most applications of
aerodynamics.} Note that the gradient of $p_t$ is a contravariant
vector field, not to be confused with a covariant derivative of
$p_t$, something which would add a \emph{covariant} index.

The reader will find in many texts that the Euler equations are
expressed without explicit use of the time subscript, and with $(U
\cdot \nabla) U$ or $U \cdot \nabla U$ substituted in place of
$\nabla_{U}U$. In terms of Cartesian coordinates, one can think of
$U \cdot \nabla$ as the differential operator obtained by taking
the inner product of the vector $U$ with the differential operator
$\partial/\partial x_1 +
\partial/\partial x_2 + \partial/\partial x_3$. In component
terms, one can write:

$$((U \cdot \nabla)U)^i = U^1 \frac{\partial U^i}{\partial x_1} + U^2
\frac{\partial U^i}{\partial x_2} + U^3 \frac{\partial
U^i}{\partial x_3} \;.$$

The \emph{streamlines} of a fluid at time $t$ are the integral
curves of the vector field $U_t$.\footnote{An integral curve
$\alpha: I \rightarrow \Omega$ of a vector field $V$ is such that
its tangent vector $\alpha'(s)$ equals the vector field
$V_{\alpha(s)}$ for all $s \in I$, where $I$ is an open interval
of the real line.} Given that $U_t$ is capable of being
time-dependent, the streamlines are also capable of being
time-dependent. The \emph{trajectory} of a particle over time can
be represented by a curve in the fluid, $\phi: I \rightarrow
\Omega$, which is such that its tangent vector $\phi'(t)$ equals
the fluid vector field $U_{(\phi(t),t)}$ at position $\phi(t)$ at
time $t$. The trajectory of the particle which is at $x \in
\Omega$ at time $t = 0$ can be denoted as $\phi_x(t) = \phi(x,t)$.
The trajectories $\phi_x(t)$ followed by particles in the fluid
only correspond to streamlines in the special case of steady-state
fluid flow, where $\partial U_t/\partial t = 0$. The mapping
$\phi: \Omega \times I \rightarrow \Omega$ enables one to define a
time-dependent set of mappings $\phi_t: \Omega \rightarrow
\Omega$. Given any volume $W$ at time $0$, $W_t = \phi_t(W)$ is
the volume occupied at time $t$ by the collection of particles
which occupied $W$ at time $0$. $\phi_t$ is called the fluid flow
map, and a fluid is defined to be \emph{incompressible} if the
volume of any region $W \subset \Omega$ is constant in time under
the fluid flow map, (Chorin and Marsden 1997, p10). A fluid is
defined to be \emph{homogeneous} if $\rho$ is constant in space,
hence a fluid which is both homogeneous and incompressible must be
such that $\rho$ is constant in space and time.

The Euler equations are derived from the application of Newton's
second law to a continuous medium, i.e., the force upon a fluid
volume must equal the rate of momentum transfer through that
volume. From the conservation of mass, one can derive the
so-called continuity equation:

$$
\frac{\partial \rho_t}{\partial t} = \text{div} \: (\rho_t U_t)
\;.
$$ In the case of an incompressible fluid, $\partial
\rho_t/\partial t = 0$ and $\text{div} \:(\rho_t U_t) = \rho \;
\text{div} \; U_t$, and, assuming $\rho > 0$, the continuity
equation reduces to:

$$
\text{div} \; U_t = 0 \;.
$$

In the more general case of a non-perfect fluid, the Cauchy stress
tensor takes the form (Chorin and Marsden 1997, p33):

$$
\sigma = \lambda(\text{div}\; U)I + 2 \mu D \;,
$$ where $\mu$ and $\lambda$ are constants, and $D$ is the
`deformation' tensor, or `rate of strain' tensor, the
symmetrisation of the covariant derivative of the velocity vector
field, $D = \text{Sym}\; (\nabla U)$. In the case of Cartesian
coordinates in Euclidean space, the components of the covariant
derivative correspond to partial derivatives, so $U^i_{;j} =
\partial U^i/\partial x_j$, and the components of the deformation
tensor are $D_{ij} = 1/2(\partial U^i/\partial x_j + \partial
U^j/\partial x_i)$.

The relevant equations for a fluid with such a Cauchy stress
tensor, derived again from Newton's second law, are the
Navier-Stokes equations. In the case of an incompressible
homogeneous fluid,\footnote{Incompressible flow is a good
approximation for aerodynamics at Mach numbers $M < 0.4$, where $M
= \frac{\parallel U \parallel} {c_s}$, the ratio of the flow speed
to the speed of sound $c_s$.} the Navier-Stokes equations take the
following form:

$$
\rho_t \left [\frac{\partial U_t}{\partial t} - \mu \nabla^2 U_t +
\nabla_{U_t}U_t \right] = - \text{grad} \; p_t + f_t \;.
$$ The constant $\mu$ is the coefficient of viscosity of the fluid
in question, and $\nabla^2$ is the Laplacian, also represented as
$\Delta$ in many texts. With respect to a Cartesian coordinate
system in Euclidean space, each component $U^i$ of the velocity
vector field is required to satisfy the equation

$$
\frac{\partial U^i}{\partial t} - \mu \left [ \frac{\partial^2
U^i}{\partial x^2_1} + \frac{\partial^2 U^i}{\partial x^2_2} +
\frac{\partial^2 U^i}{\partial x^2_3} \right ] + U^1
\frac{\partial U^i}{\partial x_1} + U^2 \frac{\partial
U^i}{\partial x_2} + U^3 \frac{\partial U^i}{\partial x_3}  = -
\frac{1}{\rho} \frac{\partial p}{\partial x_i} + \frac{f}{\rho}
\;.
$$ The mass density $\rho$ has been shifted here from the
left-hand side to the right-hand side of the equation, and the
time subscript has been removed to avoid clutter.

Only solutions of the Navier-Stokes equations are capable of
representing the presence of a \emph{boundary layer} in the fluid
flow adjacent to a solid body, a phenomenon of crucial importance
in aerodynamics.\label{boundary} The only internal forces in a
perfect fluid are pressure forces, and these act normally to any
surface in the fluid. In contrast, the boundary layer is a
consequence of friction and shear stresses within a fluid, and
such forces act tangentially to the surfaces which separate
adjacent layers in a fluid. Due to friction between a fluid and
the surface of a solid body, and the presence of shear stresses
within the fluid, there is a region of fluid adjacent to the solid
body in which the velocity is less than the free-flow velocity,
and this region is the boundary layer. The fluid velocity in the
boundary layer tends towards zero approaching the surface of the
solid body. Fluids with different coefficients of viscosity $\mu$
have boundary layers of different thickness, and different
velocity gradients within the boundary layer.

Note that the Navier-Stokes equations, and fluid mechanics
\textit{in toto}, merely provide a phenomenological approximation;
a fluid is not a continuous medium, but a very large collection of
discrete molecules, interacting with each other. Hence, fluid
mechanics is not a fundamental physical theory. As a consequence,
there is no sense in which a body of air \emph{exactly} realises a
solution of the Navier-Stokes equations. However, despite the fact
that the Navier-Stokes equations are only considered to be
phenomenological, there are notable controversies and difficulties
in using these equations to explain the most important aerodynamic
phenomenon of all: the lift generated by a wing.

\section{Bernoulli vs. Coanda}

The simplest and most popular explanations of aerodynamic lift
invoke the Bernoulli principle, which, in turn, is derived from
Bernoulli's theorem. The general version of Bernoulli's theorem
states that the following quantity,

$$
\frac{1}{2} \parallel U \parallel ^2 + w + \frac{p}{\rho} + \phi
\;,
$$ is constant along the streamlines of a stationary,
compressible, perfect isentropic fluid, assuming $p$ is a function
of $\rho$, (Abraham, Marsden and Ratiu (1988), p593). The
gravitational potential for an external gravitational field is
denoted here as $\phi$. The scalar field $w$ is the internal
energy density per unit mass of the fluid,\footnote{This is
related to the thermodynamic quantity of enthalpy per unit mass
$\epsilon$ by the equation $\epsilon = w + p/\rho$.} and an
isentropic fluid is one for which the internal energy density $w$
depends only upon the mass density $\rho$. i.e., if the fluid
undergoes compression, the internal energy per unit mass may
increase. Internal energy density is to be distinguished from the
kinetic energy density of the fluid, $1/2 \: \rho
\parallel U \parallel ^2$. If the internal energy density depends only upon
the mass density, then this rules out a change of internal energy
density due to heat transfer between the fluid and its
environment, or a change of internal energy density due to the
generation of friction.

If one makes the same assumptions, but with the exception that the
fluid is both incompressible, and homogeneous throughout space,
then the fluid will be a constant density $\rho_0$ throughout
space and time; the internal energy density will be constant $w_0
= w(\rho_0)$ also; and the following quantity will be constant
along fluid streamlines:

$$
\frac{1}{2} \parallel U \parallel ^2 + \frac{p}{\rho_0} + \phi \;.
$$ The general version of Bernoulli's theorem can be derived from the Euler
equations,

$$
\rho \left[\frac{\partial U}{\partial t} + \nabla_{U}U \right] = -
\text{grad} \; p + f \; .$$ To provide a simple derivation, let us
assume the case of an incompressible and homogeneous fluid.
Amongst other things, this entails that $\partial U/\partial t =
0$, hence Euler's equations reduce to:

$$
\rho_0 (\nabla_{U}U) = - \text{grad} \; p + f \; .$$ We shall
additionally assume that the fluid is irrotational. The covariant
derivative can be expressed as follows,

$$
\nabla_{U}U = \text{grad} \; \frac{\parallel U \parallel^2}{2} +
(\text{curl} \; U) \wedge U \; ,
$$ and an irrotational fluid is such that $\text{curl} \;
U = 0$, hence Euler's equations further reduces to

$$
\rho_0 (\text{grad} \; \frac{\parallel U \parallel ^2}{2}) = -
\text{grad} \; p + f \; .$$ Dividing both sides by $\rho_0$ one
obtains

$$
\text{grad} \; \frac{\parallel U \parallel^2}{2} = - \text{grad}
\; \frac{p}{\rho_0} + \frac{f}{\rho_0} \; ,$$ and thence

$$
\text{grad} \; \left( \frac{\parallel U \parallel^2}{2} +
\frac{p}{\rho_0} \right) = \frac{f}{\rho_0} \; ,$$ and

$$
\text{grad} \; \left( \frac{\parallel U \parallel^2}{2} +
\frac{p}{\rho_0} + \kappa \right) = \frac{f}{\rho_0} \; ,$$ for
any constant $\kappa$. Whilst $f$ specifies the external force per
unit volume, the ratio $f/\rho_0$ specifies the external force per
unit mass, which is related to an external potential $\phi$ by
$f/\rho_0 = - \text{grad}\; \phi$. For any constant $c$ it follows
that

$$
- \text{grad}\;(\phi + c) = \frac{f}{\rho_0} \;.
$$ Hence,

$$
\text{grad} \; \left( \frac{\parallel U \parallel^2}{2} +
\frac{p}{\rho_0} + \kappa \right) =  - \text{grad}\;(\phi + c) \;
.$$ From this it follows that

$$
 \frac{\parallel U \parallel^2}{2} + \frac{p}{\rho_0} + \kappa  =
-\phi - c \; ,$$ and

$$
\frac{\parallel U \parallel^2}{2} + \frac{p}{\rho_0} + \phi =
-(\kappa + c) \; ,$$ which is our desired result because $-
(\kappa + c)$ is a constant.

From Bernoulli's theorem, then, comes Bernoulli's principle, which
states that an increase in fluid velocity corresponds to a
decrease in pressure, and a decrease in fluid velocity corresponds
to an increase in pressure. Whilst Bernoulli's principle is often
invoked to explain the phenomenon of aerodynamic lift generated by
the air flow around a wing profile, there are alternative
explanations which employ, in some combination: the `Coanda
effect', the notion of circulation, and Newton's third law. These
alternative explanations are, at the very least, equally
legitimate to the Bernoulli-principle explanation, and, amongst
aerodynamicists, are considered to be superior to the
Bernoulli-explanation. There is also a long-standing popular
misconception associated with the Bernoulli-principle explanation,
which has been widely disseminated, but which is completely false.

The Bernoulli-principle approach explains the lift generated by a
wing as the consequence of lower pressure above the wing than
below, resulting in a net upward force upon the wing. The lower
pressure above the wing corresponds to faster air flow above the
wing, in accordance with the Bernoulli principle. The Bernoulli
principle itself merely states that lower pressure corresponds to
faster airflow, and \textit{vice versa}; it does not state that
faster airflow \emph{causes} lower pressure, any more than lower
pressure causes faster airflow. Faster airflow does not have
causal priority over lower pressure. However, at this point, the
misconceived popular explanation implicitly assigns causal
priority to faster airflow, explains the lower pressure as a
consequence of the faster airflow, and explains the faster airflow
above the wing as a consequence of the fact that wings have
greater curvature above than below, and the air flowing over the
top therefore follows a longer path than the air flowing below.
Conjoined with this is a curious argument, which invokes the
notion of a `packet' of air, and claims that a packet of air which
is divided at the leading edge of a wing must rejoin at the
trailing edge, hence the air traversing the longer path over the
top must travel faster to rejoin its counterpart at the trailing
edge. This `path-length' explanation of lift does not require the
air at the trailing edge of the wing to be deflected downwards,
despite the downdraught which can be experienced directly beneath
a passing aircraft. If this explanation were true, and the cause
of lift was merely the asymmetrical profiles of the upper and
lower surfaces of a wing, then it would not be possible for an
aeroplane to fly upside-down. In addition, experiment demonstrates
that air passing over the upper surface of a wing travels at such
speed that packets of air divided at the leading edge of the wing
fail to rejoin at the trailing edge.

One alternative explanation argues that wings generate lift
because they deflect air downwards, and, by Newton's third law, an
action causes an equal and opposite reaction, hence the wing is
forced upwards. According to this explanation, the trailing edge
of a wing must point diagonally downwards to generate lift, and
this is achieved either by tilting the wing downwards with respect
to the flow of air, or by making the wing cambered, or both. Air
is deflected downwards by both the lower and upper surfaces of the
wing. The lower surface deflects air downwards in a
straightforward fashion, given that the wing is either tilted with
respect to the direction of airflow, or the lower surface is of a
concave shape. However, the majority of the lift is generated by
the downwards deflection of the air which flows \emph{above} the
wing. This explanation then depends upon the Coanda effect, the
tendency of a stream of fluid to follow the contours of a convex
surface rather than continue moving in a straight line. The flow
over the surface of a wing is said to remain `attached' to the
wing surface. An aircraft wing is said to `stall' when the
boundary layer on the upper surface `detaches' or `separates', and
is no longer guided downwards by the contours of the upper
surface. In this circumstance, the lifting force generated by the
upper surface of the wing suddenly becomes very small, and the
lifting force that remains is generally insufficient to support
the weight of the aircraft. To create lift at low speed, an
aircraft must increase the angle of attack\footnote{The angle of
attack is the angle subtended from the direction of airflow by the
straight line which joins the leading edge and trailing edge of
the wing. This line is called the chordline, and its length is
called the chord of the wing.} of the wing, and the upper surface
is carefully contoured to prevent flow detachment under these
conditions.

As stated in Section \ref{boundary}, whilst the region of fluid
flow outside a boundary layer can be idealized as a perfect fluid
and represented by a solution of the Euler equations,
representation of the boundary layer itself requires the
Navier-Stokes equations. In particular, the inverse relationship
between velocity and pressure, encapsulated in Bernoulli's
principle, is only valid outside the boundary layer. Thus, despite
the velocity gradient across the boundary layer, normal to the
surface of the solid body, there is no corresponding pressure
gradient. The pressure across the boundary layer, normal to the
surface of the solid body, is represented to be almost equal to
the pressure just outside the boundary layer. The pressure inside
the boundary layer changes along its length as the pressure
outside the layer changes. Towards the trailing edge of a wing,
the pressure outside the boundary layer builds, and the velocity
outside the layer decreases, approaching zero towards the aft
`stagnation point'. This has the consequence that the range of
fluid flow velocities inside the boundary layer also decreases. If
the pressure increase towards the stagnation point is not
sufficiently gradual, then the boundary layer can separate from
the surface of the wing before the trailing edge. This occurs
because the velocities inside the boundary layer are already
smaller than those outside, hence they can reach zero before the
aft stagnation point. After this, the velocities inside the
boundary layer can reverse direction, and this causes the flow to
separate from the wing. Irrespective of where the boundary layer
separates, it then breaks up into eddies and forms a turbulent
wake behind the wing.

The boundary layer airflow which remains attached to the upper
surface of the wing, does so only because the pressure outside the
boundary layer is slightly higher than the pressure inside the
boundary layer, so there is a pressure gradient which forces the
boundary layer to \emph{apparently} adhere to the convex upper
surface of the wing. There is no genuine force of attraction
between the wing surface and the boundary layer airflow.

The lift generated by a wing is also often explained to be a
consequence of non-zero circulation in the airflow around the
wing. Given a closed loop $C$ parameterized by $s$, in a fluid
with a velocity vector field $U$, the circulation around $C$ is
defined by the line integral

$$
\Gamma_C = \oint_C \langle U , d/ds \rangle \: ds \;,
$$ where $d/ds$ is the tangent vector to the loop
$C$. To understand how lift can be explained by non-zero
circulation, consider the idealized situation where the airflow
around a wing is a superposition of a uniform `freestream' flow
from left to right, where the streamlines are parallel straight
lines, and a pure circulatory flow, where the streamlines are
clockwise concentric circles. Taking the sum of the velocity
vector fields for the uniform and circulatory flow at each point,
the clockwise circulation is added to the freestream velocity
above the wing, and subtracted from it below. The Bernoulli
principle can then be invoked to explain the pressure differential
above and below the wing, which, in turn, explains the net upward
force. The presence of circulation also accords with the use of
Newton's third law to explain the uplift, because the circulatory
flow adds a downward component to the airflow in the wake of the
wing, and this corresponds to the downward deflection of air.

The notion of circulation can be used to quantitatively explain
the lifting force on a wing by the Kutta-Joukowski theorem. If $C$
is a loop enclosing a wing profile, and if an incompressible
airflow of density $\rho$ has a velocity vector field which tends
towards a constant $U$ as the distance from the wing tends to
infinity, then the Kutta-Joukowski theorem asserts that the
lifting force per unit span\footnote{The span is the width of a
wing in a direction transverse to the airflow.} on a notional wing
of infinite span, is given by the equation

$$L=-\rho \: \Gamma_C
\parallel U \parallel \textbf{n} \;,$$ where $\textbf{n}$ is a unit vector
orthogonal to $U$, (Chorin and Marsden 1997, p53). In reality, the
lifting force decreases towards the tips of a wing, hence the
simplifying assumption here of an infinite wing span, which allows
one to treat the flow as if it were two-dimensional. The unit
vector $\textbf{n}$ is orthogonal to $U$ in a two-dimensional
plane which represents a longitudinal cross-section of the flow
around the wing.

The presence of a circular component to the airflow around a wing
profile cannot be explained without introducing the notions of
viscosity, the boundary layer, the Coanda effect, and boundary
layer separation. The absence of circulation requires the boundary
layer to detach forward of the trailing edge upon the upper
surface of the wing. Upon the commencement of airflow, such
premature detachment does indeed occur, but this is a transient
and unstable situation, which, assuming left to right airflow,
generates anti-clockwise vortices, called starting vortices, until
the boundary layer separation point migrates to the trailing edge
of the wing. Once the airflow above and below the wing is
detaching from the trailing edge, which is directed diagonally
downward, there is a non-zero clockwise circulatory component to
the airflow.

The highest airflow velocity and lowest pressure occurs towards
the front of the upper surface of a wing. The presence of
circulation raises the fluid velocity above the wing, and the low
radius of curvature at the front of the wing accentuates the
effect. This is because the velocity of fluid in circulation is
function of the radius of curvature. With the exception of very
low radii, the velocity of circulating fluid is inversely
proportional to the radius. Hence, where the radius of curvature
of a wing is lowest, around the nose of the wing, the fluid flow
is accelerated to its highest velocity, and the pressure of the
fluid drops to its minimum. This effect is accentuated above the
wing because of the direction of the circulation. The radius of
curvature increases towards the rear of the wing, hence the
velocity decreases and the pressure increases towards the trailing
edge of the upper surface.

Both the Bernoulli-principle explanation and the circulation
explanation of lift fall under the aegis of the
deductive-nomological (DN) account of scientific explanation. In
such explanations, one explains certain phenomenal facts by
logically deriving them from the conjunction of general laws and
particular specified circumstances. As Torretti emphasises (1990,
p22-24), such explanations in physics typically require a
re-conceptualization of the phenomenal facts in the same terms in
which the laws and circumstances are stated. A fluid mechanics
explanation of the lift generated by a wing represents the wing as
a compact subset $M$ of Euclidean space, which is surrounded by a
continuous medium that possesses a fluid flow vector field, a
pressure scalar field, and a mass density scalar field satisfying
the Navier-Stokes equations, and represents the presence of lift
by a non-zero vertical component to the vector which represents
the net force upon the wing. The Bernoulli-principle explanation
represents the net force upon the wing as the integral
$-\int_{\partial M} p \textbf{n}$ of the pressure force over the
surface $\partial M$ of the wing, where $\textbf{n}$ is an
outward-pointing unit normal; the circulation explanation
represents the vertical force per unit span as $-\rho \:
\Gamma_{C}
\parallel U \parallel \textbf{n} = -\rho \: [\oint_{C} \langle U , d/ds \rangle \:
ds] \parallel U \parallel \textbf{n}$, where $C$ is a loop
enclosing the longitudinal profile of $M$, and $\textbf{n}$ is a
unit vector orthogonal to $U$.\footnote{Here we have neglected the
fact that the lifting force per unit span decreases close to the
wing tips.}

Although the Bernoulli principle can be used to determine the net
force and the lifting force upon a solid body once the pressure
and velocity distribution around the body have been specified,
this flow regime is a consequence of viscosity, and cannot be
explained without recourse to the Coanda effect, circulation, and
the curvature of the wing. Hence the Bernoulli principle
explanation is subservient to an explanation of lift which employs
the latter concepts.

Whilst the debate over the appropriate explanation for aerodynamic
lift is a debate over the appropriate fluid mechanical explanation
for lift, fluid mechanics is, of course, a continuum idealization,
and, in reality, aerodynamic lift is the consequence of the
aggregate effect of the interactions between the molecules in the
air flow and the molecules in the surface of the wing. Charles
Crummer explains the Coanda effect in these terms as follows:

``In a steady flow over a surface, stream particles have only
thermal velocity components normal to the surface. If the surface
is flat, the particles that collide with boundary layer particles
are as likely to knock them out of the boundary layer as to knock
others in, i.e. the boundary layer population is not changed and
the pressure on the surface is the same as if there were no flow.
If, however, the surface is curved in a convex shape, the
particles in the flow will tend to take directions tangent to the
surface, i.e. away from the surface, obeying Newton's first law.
As these particles flow away from the surface, their collisions
with the boundary layer thermal particles tend to ``blow" those
particles away from the surface. What this means is that if all
impact parameters are equally likely, there are more ways a
collision can result in a depletion of the boundary layer than an
increase in the boundary layer population. The boundary layer will
tend to increase in thickness and to depopulate; the pressure will
reduce there. Those particles in the flow that do interact with
the stagnant boundary layer will give some of their energy to
particles there. As they are deflected back into the flow by
collisions with boundary layer particles, they are, in turn,
struck by faster particles in the flow and struck at positive
impact parameters. They are thus forced back toward the surface.
This is how the flow is ``attracted" to the surface, the Coanda
effect," (Crummer 2005, p8).

\section{Case Study: Aerodynamics in Formula 1}

Formula 1 is an economic, technological and sporting activity, in
which an extended physical object, the car, is equipped with an
internal combustion engine, capable of converting the chemical
potential energy from a supply of fossil fuel into useful work. A
Formula 1 car is almost always in contact with a ground plane, and
this distinguishes the nature of the airflow from that experienced
by an aircraft in free flight.

As Peter Wright points out, ``A Formula 1 car can be described
aerodynamically as a very low aspect ratio\footnote{The aspect
ratio for a wing of span $s$ and area $S$ is $s^2/S$; for a
rectangular wing, the aspect ratio is the ratio of the span to the
chord, (Katz 1995, p115).} (0.38) bluff body in close proximity to
the ground (gap/chord = 0.005). It is surrounded by low aspect
ratio rotating cylinders in contact with the ground, and it is
equipped with leading and trailing edge airfoils, the leading edge
device also being in close proximity to the ground. There are
major internal flows, with heat addition (water, oil, and brake
cooling), an air intake, hot gas injection, and a large open
cavity (the cockpit). The flow is three-dimensional and has
extensive areas of separated flow and vorticity," (2001, p125).

The objective of a racing car designer is to create a car which,
under the control of a human being, traverses a range of closed
circuits at the greatest average speed possible. This requires a
car which accelerates, brakes, and corners as quickly as possible,
within the constraints established by the technical and sporting
regulations. In physical terms, one wishes to maximize the
longitudinal and lateral acceleration which the car is capable of
generating. The tyres of a car provide the contact surfaces
between the car and the road surface, and the frictional grip
provided by the tyres determines the upper limit on the
longitudinal and lateral forces which the car is capable of
generating.

The grip generated by a tyre is a function of the force pushing
that tyre onto the road surface. The maximum horizontal force
$F_h$ which a tyre can generate is $C_f F_v$, where $C_f$ is the
tyre's peak coefficient of friction, and $F_v$ is the peak
downward vertical force on the tyre. As a consequence, the greater
the weight of a car, the greater the grip generated by its tyres.
However, greater weight generally fails to increase the
longitudinal and lateral acceleration of a car because the
increase in the longitudinal and lateral force which the car is
capable of generating is cancelled out by the greater inertia
which is a concomitant of the greater weight.

Two aerodynamic revolutions can be identified in the history of
Formula 1: the discovery of `downforce' by means of inverted wing
profiles, and the discovery of ground effect downforce. As a
consequence of the first revolution, Formula 1 designers are able
to generate an aerodynamic force which is directed downwards
(`downforce') by attaching appendages to the car which are, in
effect, inverted versions of the wing profiles on aircraft. This
increases the force pushing the tyres onto the road surface
without any increase in inertia. Although downforce brings an
inevitable drag force penalty with it, the increase in the lateral
accelerative capabilities of a Formula 1 car increases the average
speed of a Formula 1 car over most types of closed circuit. Both
the drag force and the downforce are proportional to the square of
the velocity of a car. The drag force is given by

$$
F_{drag} \approx \frac{1}{2} C_D A \rho v^2 \;,
$$ where $C_D$ is the drag coefficient, a dimensionless number determined by the exact
shape of the car and its angle of attack; $A$ is the frontal area;
$\rho$ is the density of air; and $v$ is the speed of the car. The
downforce is given by

$$
F_{downforce} \approx \frac{1}{2} C_L A \rho v^2 \;,
$$ where $C_L$ is the coefficient of lift, again determined by the
exact shape of the car and its angle of attack. In the case of a
modern Formula 1 car, the lift-to-drag ratio $C_L/C_D$ has a
typical value of, say, $2.5$ (Wright 2001, p125), so downforce
dominates performance.

\begin{figure}[h]
\centering
\includegraphics[scale = 0.3]{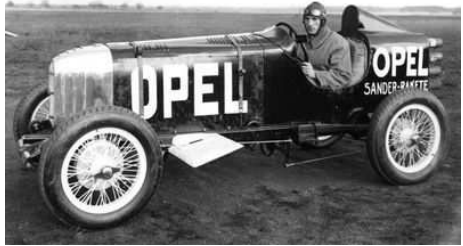}
\caption{Opel's 1928 experimental rocket car, the RAK 1, with
side-wings.} \label{Rak1}
\end{figure}

\begin{figure}[h]
\centering
\includegraphics[scale = 0.4]{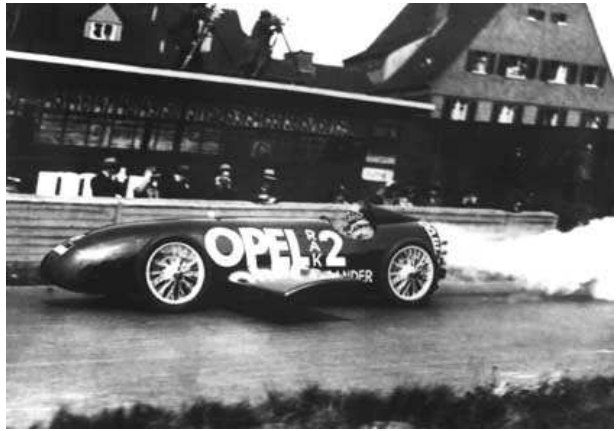}
\caption{Opel's RAK 2, with enlarged side-wings.} \label{Rak2}
\end{figure}

\begin{figure}[h]
\centering
\includegraphics[scale = 0.4]{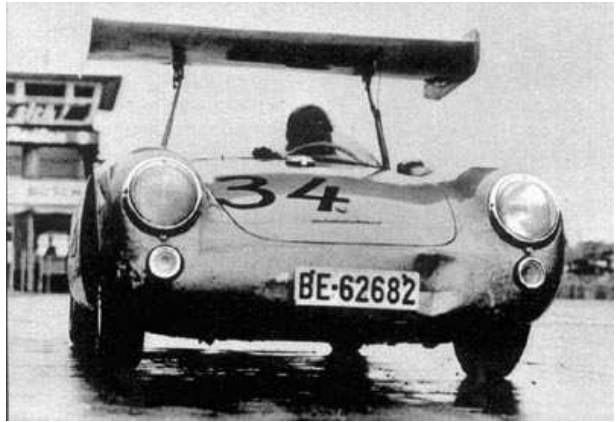}
\caption{Michael May's 1956 Porsche 550 Spyder, with inverted
wing.} \label{May}
\end{figure}

Inverted wing profiles were introduced to Formula 1 rather
belatedly, after a process of gradual discovery occurring over
many decades, and involving multiple individuals within the
community of automotive engineers, (Hughes 2005, p119-121; Katz
1995, p245-250). In fact, the idea of aerodynamic downforce has a
provenance which can be traced back to 1928. Inverted wing
profiles were initially used to provide stability or to counteract
the tendency a car might have to lift at high speeds. In 1928,
Opel's experimental rocket-powered cars, the RAK 1 and RAK 2, were
equipped with inverted wings on either side between the wheels to
counteract high-speed lift (see Figures \ref{Rak1} and
\ref{Rak2}). In 1956 a Swiss engineer and amateur racing driver
called Michael May experimented with an inverted wing mounted over
the cockpit of his Porsche 550 Spyder (see Figure \ref{May}).
Michael and his brother Pierre had recalled the use of such wings
upon the Opel RAK. When May's car proved faster than the works
Porsches, Porsche lobbied successfully for the appendage to be
banned, under the pretext that it obscured the vision of following
drivers, and May failed to pursue the idea any further.

\begin{figure}[h]
\centering
\includegraphics[scale = 0.5]{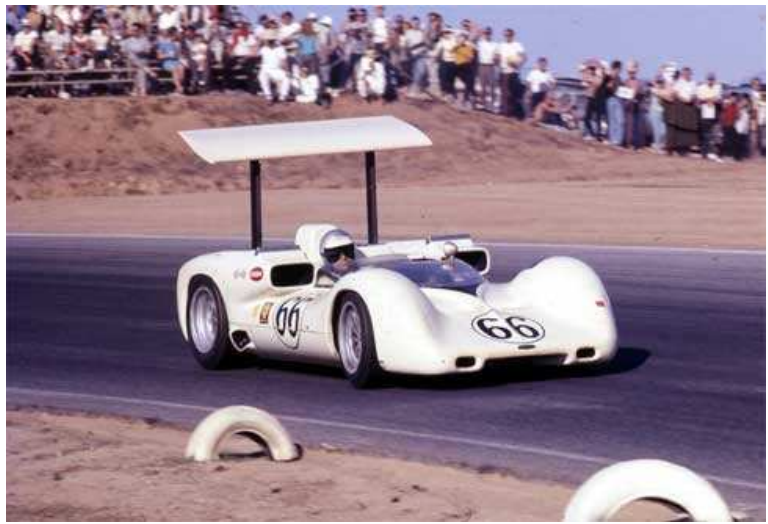}
\caption{The high-winged Chaparral 2E of 1966.}
\label{Chap}
\end{figure}

Perhaps the most influential innovator in the field of racing car
aerodynamics was Texan oil magnate, engineer and driver Jim Hall.
Hall's Chaparral company linked up with the Chevrolet R\&D
department, headed by Frank Winchell, and competed in US sportscar
championships such as Can-Am in the 1960s. In 1961, the Chaparral
1 sports car experienced lift at high speed, and Bill Mitchell,
chief stylist of General Motors in the 1950s and 1960s, suggested
using an inverted wing (Wright 2001, p299). In 1963 protruding
front-mounted wings were fitted to prevent the front wheels of the
Chaparral 2 from lifting off the ground (Wright 2001, p300). In
1965, the Chaparral 2C was fitted with a rear wing mounted on
pivots, with a driver-adjustable angle of attack, and in 1966 the
concept was extended to a dramatic high wing on the Chaparral 2E,
(see Figure \ref{Chap}).

In 1966 the McLaren F1 team tested wings with great success, but
due to lack of resources, the team never found the time to revisit
the idea, as recounted by chief designer Robin Herd (Hughes 2005,
p120): ``We didn't want anyone else to twig, so we took the wings
off, quietly put them in the back of the truck and continued with
our normal testing. We decided we would look into it further, in
private, when we had the time. But an F1 team in those days was so
madly understaffed that we never got round to looking at it
properly."

\begin{figure}[h]
\centering
\includegraphics[scale = 0.5]{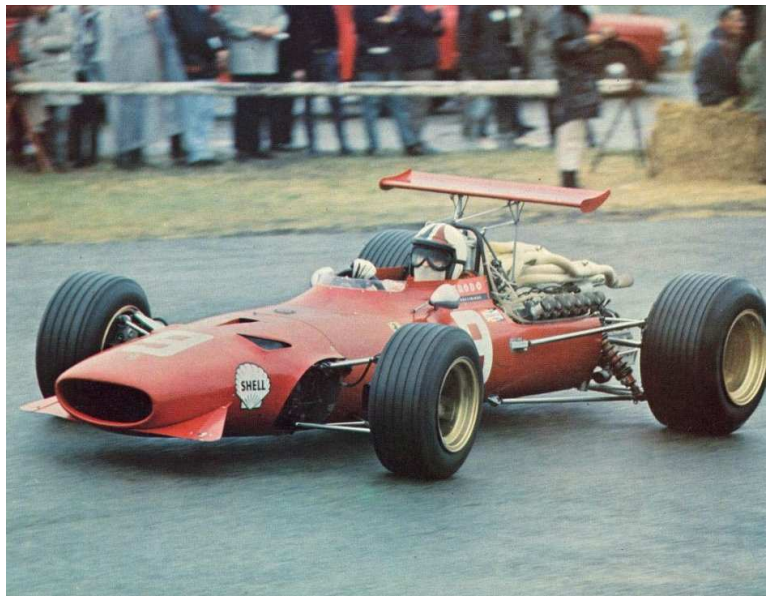}
\caption{The winged Ferrari 312 of 1968.} \label{Ferrari}
\end{figure}

In November 1967, Jim Clark raced an American Indycar called a
Vollstedt, which possesed small wings, and during the Tasman
winter series Clark recounted the grip and stability produced by
this car to one of his Lotus mechanics. The mechanic extemporised
a make-do wing from a helicopter rotor, which was fitted to
Clark's car, and swiftly removed, but not before a Ferrari
engineer had taken photographs of it. At the 1968 Belgian Grand
Prix, Ferrari appeared with full inverted rear wings (see Figure
\ref{Ferrari}), and Brabham did likewise on the day after
Ferrari's wings first appeared. As Mark Hughes recounts,
``Ferrari's chief engineer Mauro Forghieri, his memory perhaps
triggered by that Tasman photo taken by one of his staff of that
experimental Lotus wing, had recalled that Michael May - the
engineer with whom he had worked perfecting Ferrari's fuel
injection system a few years earlier - had once made a wing for
that sports car of his. `Michael was a friend as well as a
consultant,' says Forghieri. `He told me about the improvement in
handling of his winged Porsche. The Chaparral convinced us even
more about the idea'," (Hughes 2005, p121). Given that May's idea
was, in turn, inspired by the Opel RAK, the introduction of
downforce to Formula 1 in 1968 can be traced back to Opel's
experimental rocket car of 1928!

The utility of downforce seems so obvious in retrospect, that its
rather belated application in Formula 1 seems puzzling. The
stories told about scientific discovery often seem to involve
either a deliberate, conscious, trend of thought that finally
reaches some goal, or sudden leaps of the imagination, or
serendipitous discoveries made by minds receptive to new ideas. In
the case of aerodynamic downforce, an example of engineering
discovery, none of these descriptions seem to apply. Seeking to
explain this, Peter Wright asserts that ``the possible magnitude
of the downforce that could be generated was not appreciated,
despite all the data being available within the aeronautical
industry. It was probably also considered that the tires
(\textit{sic}) would not be able to sustain the additional work
and that the drag penalty associated with generating the downforce
would result in a net loss of average speed...It may also have
been simply an attitude among designers, most of whom came from
production car backgrounds-the cars did not have wings. It took
the free spirits of Frank Winchell and Jim Hall, and their
measurement and simulation groups at Chevrolet R\&D and Chaparral,
to cast aside convention and explore the possibilities of this new
and now obvious way to increase performance," (Wright 2001, p124).
The reluctant adoption of downforce in Formula 1 seems to be an
example of the trammelling effect of convention in academic and
research communities. It is a demonstration that the significance
of new empirical facts often go unrecognised when they are not
subsumed within the appropriate conceptual scheme: aerodynamics
were only considered to be relevant to Formula 1 to the extent
that they enabled a designer to minimise drag. One might say that
a particular engineering paradigm existed in Formula 1,
unquestioned, until the late 1960s, when the amount of suggestive
empirical data built to such an extent that a revolution was
unavoidable.

\begin{figure}[h]
\centering
\includegraphics[scale = 0.5]{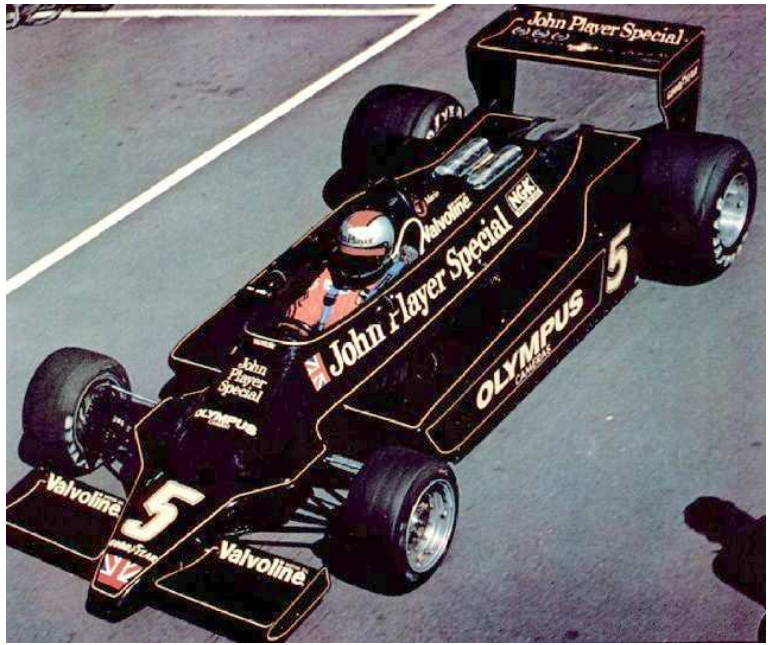}
\caption{The skirted ground effect Lotus T79 of 1978.}
\label{Lotus}
\end{figure}

The second revolution in Formula 1 aerodynamics occurred
approximately a decade after the first, with the introduction of
the Lotus T78 in 1977, and its further development, the Lotus T79
in 1978 (see Figure \ref{Lotus}). It was discovered that large
amounts of downforce could be generated from the airflow between
the underbody of the car and the ground plane. In particular, low
pressure could be created underneath the car by using the ground
plane almost like the floor of a venturi duct. The ceiling of
these venturi ducts took the form of inverted wing profiles
mounted in sidepods between the wheels of the car. The decreasing
cross-sectional area in the throat of these ducts, and the
inverted wing profile accelerated the airflow and created low
pressure in accordance with the Bernoulli principle. The gap
between the bottom of the sidepods and the ground was sealed by
so-called `skirts'. When the rules permitted it, the skirts were
suspended from the sidepods with a vertical degree of freedom
(Katz 1995, p200) to maintain a constant seal with the ground
under changes in the attitude and ride height of the car.

Brabham responded to the Lotus `wing car' with the BT46B `fan
car', in which low pressure was created under the car by a
rear-mounted fan driven off the gearbox. Such a car is able to
generate downforce independently of the speed of airflow under the
car, and the BT46B won its debut race in 1978, and was promptly
banned by the governing body.

Whilst the inverted wing profiles in the sidepods of ground effect
cars were often dubbed `sidewings', and the bottom of the car was
often dubbed the `underwing', these names are potentially
misleading because the downforce generated is largely dependent
upon the presence of the ground plane, and would not be generated
in free flow. Note also that the tunnels created under the car are
not strictly venturi ducts either. In a venturi duct, the sides of
the duct do not move with respect to each other. In contrast,
underneath such a car the ground plane and the airflow are both
moving with respect to the walls and ceiling of the ducts.

Again, this second revolution in Formula 1 aerodynamics could have
been implemented many years before it was. Jim Hall and Chevrolet
R\&D had tried shaping the entire underbody as an inverted wing
profile in the 1960s, and the 1970 Chaparral 2J had been equipped
with two rear-mounted fans to lower the air pressure under the
car, and flexible plastic skirts to seal the low pressure area.
The Chaparral 2J was banned by the governing body of Can-Am at the
end of the season under lobbying from fellow competitors. In 1969,
Formula 1 engineer Peter Wright designed an inverted wing body
shape for BRM which yielded promising results in the Imperial
College wind-tunnel, (Wright 2001, p301). However, the design
never saw the light of day due to a management upheaval. Peter
Wright joined Lotus some years later, and made a serendipitous
discovery in the Imperial College wind-tunnel in 1975: ``By the
end of a week of tunnel testing, the strong wooden model would
have been so modified with card, modeling clay, and sticky
tape...that it had usually lost most of its structural integrity.
Toward the end of one of the weeks in the tunnel, I noticed that
it was becoming almost impossible to obtain consistent balance
readings. Something was wrong. Looking carefully at the model, it
became clear that the side pods were sagging under load and that
as the speed of the tunnel increased, they sagged even more. That
indicated two things: (1) that the side pods had started to
generate downforce, and (2) that it had something to do with the
gap between their edges and the ground.

``Thin wire supports restored the side pods to their correct
position and stopped them from sagging - no downforce and
consistent balance readings. Next we taped card skirts to seal the
gap between the edge of the side pods and the ground, leaving only
approximately 1mm (0.04in) gap. The total downforce on the car
doubled for only a small increase in drag!" (ibid., p302).

From 1983 onwards, the technical regulations in Formula 1 required
the cars to have flat bottoms with no skirts, and from mid-1994
onwards, the cars were required to have flat bottoms with a
central step running the length of the underbody, yet substantial
ground effect downforce could still be generated by accelerating
the airflow under the car. A flat plane inclined downwards with
respect to the airflow will generate downforce, and this effect
will be amplified by ground effect. In addition, more downforce
can be generated by using an upswept `diffuser' between the wheels
at the rear of the car to force air upwards by means of the Coanda
effect.

The Coanda effect is also utilised on a modern Formula 1 car with
the purpose, not of generating downforce directly, but of guiding
and conditioning airflow in one place, as a means of maximising
downforce elsewhere. For example, the rear of a modern Formula 1
car is tightly tapered between the rear wheels, rather like the
neck and shoulders of a coke-bottle. By means of the Coanda
effect, the air flowing along the flanks of the sidepods adheres
to the waisted contours at the rear, and the airflow here is duly
accelerated, creating lower pressure. In itself, this tranverse
pressure differential on either side of the car cancels out, and
creates no net force. However, the accelerated airflow between the
rear wheels and over the top of the diffuser does raise the
velocity of the air exiting the diffuser. The accelerated airflow
over the lower surface of the rear wing also contributes to
raising the flow rate through the diffuser, thereby enhancing the
downforce it generates. In addition, coaxing air away from the
rear tyres, and injecting high speed airflow into the region of
flow separation behind the car, both contribute to reducing drag.

\begin{figure}[h]
\centering
\includegraphics[scale = 0.5]{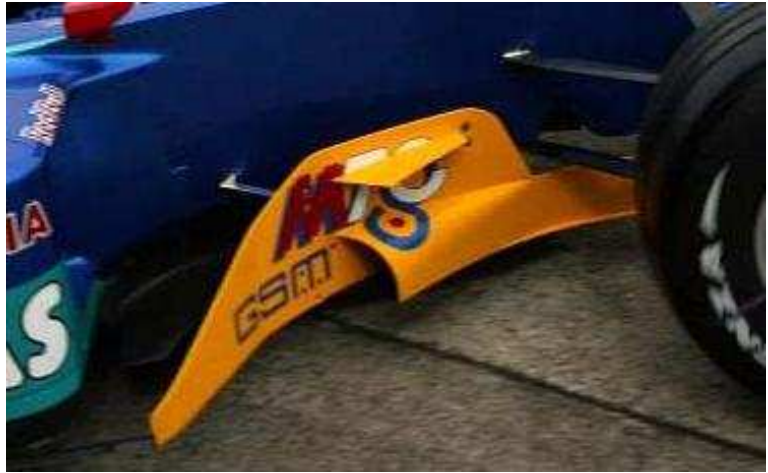}
\caption{Modern Formula 1 bargeboard.} \label{barge}
\end{figure}

The Coanda effect is also used by the so-called `bargeboards',
aerodynamic appendages typically sited between the trailing edge
of the front wheels and the leading edge of the sidepods (see
Figure \ref{barge}). Bargeboards are used to guide turbulent air
from the front wing wake, away from the vital airflow underneath
the car. In addition, the lower trailing edge of a bargeboard
creates a vortex which travels down the outer lower edge of the
sidepod, acting as a surrogate skirt, helping to seal the lower
pressure area under the car. Such techniques, and the continued
utility of ground effect in Formula 1, demonstrate the
irreversibility of conceptual revolutions, even in the case of an
engineering activity which faces increasingly constrained
technical regulations.

\end{document}